\documentclass[10pt]{iopart}

\usepackage{graphicx}
\usepackage{epsfig}
\usepackage{subfigure}
\usepackage{color}

\begin{document}

\title[Stripe phase in Bose-Einstein Condensate]{Lattice induced stripe phase in Bose-Einstein condensate under non-inertial and inertial motion}

\author{Priyam Das}
   \address{Department of Physics, Indian Institute of Technology Delhi, Hauz Khas, New Delhi - 110016, India}

\ead{daspriyam3@gmail.com}

\vspace{10pt}

\begin{abstract}
We consider a parametrically forced Bose-Einstein condensate in the combined presence of an optical lattice and harmonic oscillator potential in the mean field approach. A spatial symmetry broken Bose-condensed phase in non-inertial and inertial frame yields a stripe phase in the presence of both cubic and quintic nonlinearities. We show that existence of such stripe phase solely depends on the interplay between the quintic nonlinearity and lattice potential. Furthermore, we observe that a time-dependent harmonic oscillator frequency destroys such stripe ordering. A linear stability analysis of the obtained solution is performed and we found that the solution is stable. In order to gain better understanding of the underlying physics, we compute the energy, showing nonlinear compression of the condensate in some parameter domain.
\end{abstract}



%
%

\section{Introduction}

Bose-Einstein condensate (BEC) in a periodic potential has been an area of active research for more than two decades \cite{morsch2006,beinke2017}. It is truly an interdisciplinary field, which has connections with many areas of physics: electrons in crystal lattices \cite{lewenstein2007ultracold,jaksch2003creation}, polarons \cite{nakano2017bose,chikkatur2000}, photons in optical fibers \cite{chang2008crystallization}, gauge theories \cite{zohar2015quantum} and exotic phase transitions \cite{bloch2005ultracold,bloch2005quantum,chen2008,bloch2008}, to mention a few. A major advantage of the Bose-condensed gas in a periodic potential is its tunability over a wide range of parameter domain. For example, the depth and width of an optical lattice can easily be tuned by controlling the intensity and frequency of the laser beam respectively, whereas, the two-body atom-atom interaction strength can be controlled by means of Feshbach resonance. This has opened up the possibility to realize many novel phases of matter, existing in theoretical realm \cite{beinke2017,kartashov2011,lin2008,zhang2007band,smerzi2002,bronski2001}, many of which have been realized in experiments as well \cite{leonard2017supersolid,baumann2010dicke,fallani2004,cataliotti2003superfluid,greiner2002quantum}.

The experimental realization of spin-orbit coupled BECs has generated much attention in the present literature due to its unique characteristic in studying various phases. Among, one such phase is the supersolid phase, whose existence is characterized by the spontaneous breaking of the following two symmetries - (i) a gauge symmetry, which leads to the off-diagonal long-range order that mimics a superfluidity, and (ii) a translational symmetry, gives rise to the diagonal long range order, mimicking crystalline structure. However, another phase observed in the literature satisfying (ii) i.e., with a spontaneously broken spatially symmetric profile, is known as stripe phase or superstripes. A number of papers have already studied the properties of the stripe phase in a spin-orbit coupled BEC systems \cite{mivehvar2015,martone2014,li2013,wang2010}. In the weak coupling regime of spin-orbit coupled BEC, the resulting Hamiltonian possesses a spectrum with a doubly degenerate ground state. Therefore, the condensate can reside at each minimum of the single particle energy-momentum dispersion. The resulting interference pattern yields a stripe phase in the system. The stripe phases are also well-known in high-temperature superconductivity, where it arises due to the interplay between antiferromagnetic interactions among the magnetic ions and the Coulomb interactions between the electric charges \cite{emery1999stripe}. The stripe phase has also been predicted in Bose-Hubbard model in a triangular optical lattice \cite{wu2006quantum}, quantum hall system \cite{demler2002quantum}, core-corona system \cite{pattabhiraman2017formation} etc.

Till date, a lot of research has been carried out to investigate the dynamics of a BEC in an optical lattice, where only cubic nonlinearity is present. However, BEC with both cubic and quintic nonlinearity in the presence of an optical lattice potential has not received much attention. Quintic nonlinearity is in general considered as the phenomenological manifestation of the three-body interaction. For higher densities of the Bose-condensed gas at absolute zero, the three-body collisions play a crucial role, where it is modelled by a quintic nonlinearity in a modified Gross-Pitaevskii equation in the mean-field theory. One of the major application of quintic nonlinearity is its use in Tonks-Girardeau (TG) gas \cite{dunjko2001,paredes2004tonks,choi2015monopole}. Apart from TG gas, quintic nonlinearity appears in a number of references. Recently, the formation of breathers through spatiotemporal vortex light bullets is discussed in a cubic-quintic nonlinear medium \cite{adhikari2017}. Oscillation of a dark soliton in a BEC with both cubic and quintic nonlinearities is observed, where the quintic nonlinearity in one dimension arises due to the interplay between the radial and axial degrees of freedom \cite{anatoly2009}.  A classical dynamical phase transition in BEC occurring through the loss of superfluidity in the mean-field regime is also proposed, where both cubic and quintic nonlinearity are considered apart from an optical lattice potential \cite{das2009loss}. It is worth mentioning that such quintic term often leads to interesting phenomena through instabilities, e.g., Faraday waves \cite{abdullaev2015faraday}, Bose-Nova effect \cite{lahaye2008} etc. However, the quintic nonlinearity combining with an optical lattice potential can suppress these instabilities, which leads to many interesting physics.

Motivated by the above works, we propose a scheme to observe stripe-ordered phase in BEC, in the presence of both cubic and quintic nonlinearity. In addition to the optical lattice potential, we consider that the system is trapped in a harmonic oscillator potential as well. For the better understanding of the underlying physics, we consider all the associated parameters are time-dependent, however, this makes the system quite complicated. In order to simplify the model, we transform the equation of motion into the 	center of mass frame. We analyze the following two possible cases: (a) when the center of mass (COM) is moving with a time-dependent velocity. This is also known as the non-inertial frame, where COM is bustling with finite acceleration. It is found that the finite acceleration of the COM solely depends on the frequency of the harmonic trap. (b) When the harmonic trap is switched off, the COM is either moving with a constant velocity or becomes static, depending on the chirped profile. This is known as the inertial frame of reference, where the COM has zero acceleration.

In order to solve such highly complicated inhomogeneous system, we employ the self-similar method, also known as the F-expansion method, where the corresponding excitations are necessarily chirped. We show that a spatially broken symmetry of the condensate leads to a stripe phase in both non-inertial and inertial frame. This is solely due to the interplay between lattice potential and quintic nonlinearity. In the case of pure cubic nonlinearity, the periodicity of the lattice potential induces the same periodicity in the density modulation \cite{lin2008,zhang2007band,bronski2001}. However, the periodicity of the density modulation becomes twice to that of the lattice potential, when the quintic nonlinearity is present apart from the cubic nonlinearity. This brings a restriction on the density modulation and hence, the superfluid matter is found to exist only in certain parameter domains, yields a stripe phase. We observe that this stripe phase exists both in the presence and absence of the harmonic trap. A linear stability is performed in order to check the stability of the obtained solution and we found that the solution is stable. On the other hand, the presence of harmonic trap in the system necessitates the presence of a chirped phased, which yields an efficient nonlinear compression in some parameter regime. In order to gain a better understanding of the nonlinear compression, we compute the energy of the system. In presence of a repulsive (regular) harmonic trap, these stripe phases lead to resonances. When the frequency of the chirped pulses is in resonance with the frequency of the harmonic trap, a significant increase in kinetic energy is observed, which gives rise to the nonlinear compression of the condensate. In presence of expulsive harmonic trap (inverted), the energy initially increases and then decreases. In this case, the periodic excitations initially move towards the center of the trap and gains energy. However, due to the instabilities at the top of the expulsive trap, the excitations move downhill, thus the corresponding energy decreases. In the absence of the trap, the lattice potential plays a dominant role. When the lattice moves with a constant velocity, the energy decreases, due to the expansion of the BEC. We observe another resonance behavior when the lattice is static, since the BEC undergoes a rapid nonlinear compression. Such resonance arises when the amplitude of the sinusoidal modulation is comparable with its background density. We would like to highlight that this is analogous to the effective pulse compression in nonlinear optical fiber, observed by Moore \cite{moores1996nonlinear}.

\section{Model and equation of motion}

In this section, we briefly introduce the model and corresponding equation of motion of the system. We consider a BEC, immersed in a time modulated optical lattice potential, where both two- and three-body interactions are present. In the present scenario, an effective three-body interaction appears through a quintic nonlinearity. The nonlinear Schr\"odinger equation (NLSE) is a generic model, describing the dynamics of Bose-Einstein condensate. We would like to emphasize that quintic nonlinearity appears in nonlinear fiber optics, where the same describes the behavior of the light pulses in an optical fiber \cite{agrawal2007nonlinear}. In general, for the pico-second light pulses, the NLSE admits group velocity dispersion, mimicking an interaction, known as ``self-phase modulation". However, if one introduces a shorter pulse propagation (of the order of femto-second) by increasing the intensity of the incident light, additional nonlinear effects become important. The dynamics of such short pulse propagation can be described by a generalized NLSE, which includes higher order nonlinear terms \cite{agrawal2007nonlinear,abdullaev2005gap}. Keeping in mind the above consequences, we consider a BEC with cubic and quintic nonlinearities, loaded in a time modulated optical lattice potential. The dynamics of such a system is described by the following generalized NLSE \cite{salasnich2007matter,salasnich2002effective,utpal2010complex},
\begin{eqnarray}
i \hbar \frac{\partial \Psi}{\partial t} = \left(- \frac{\hbar^{2}}{2 m} \nabla^{2} + V_{ext} + U_{1} |\Psi|^{2} + U_{2}|\Psi|^{4} - \bar{\nu}(t)\right)\Psi.
\end{eqnarray}
where, $\Psi\equiv \Psi(x,y,z,t)$ is the condensate order parameter, $V_{ext} \equiv V_{ext}(x,y,z,t) = V_{tr}(x,y) + V_{l}(z,t)$ is the external trapping potential and $\bar{\nu}(t)$ is the time dependent chemical potential. We consider general time-dependent two-body ($U_{1} \equiv U_{1}(t)$) and three-body ($U_{2} \equiv U_{2}(t)$) interactions, which can be tuned by using Feshbach resonance technique. In order to have a cigar-shaped BEC, we apply a strong harmonic confinement along the transverse directions ($x$-$y$),  $V_{tr}(x,y) = \frac{1}{2}m \omega^{2}_{\perp} (x^{2} + y^{2})$ with a trapping frequency $\omega_{\perp}$ and harmonic oscillator length scale $a_{\perp} = \sqrt{\hbar/m \omega_{\perp}}$. Due to the symmetry of the system, we make separation of variables along the transverse and longitudinal direction $\Psi = \psi(z,t) \phi_{0}(x,y)$, where the the transverse component $\phi_{0}(x,y)$ is assumed to be a normalized Gaussian function. After a lengthy algebra, one obtains a generalized NLSE in quasi-one dimension \cite{abdullaev2005gap,choi1999bose}:
\begin{eqnarray}
i \hbar \frac{\partial \psi}{\partial t} = \left(- \frac{\hbar^{2}}{2 m} \frac{\partial{^2} }{\partial z^{2}} + V_{l} + g_{1}(t) |\psi|^{2} + g_{2}(t)|\psi|^{4} - \nu(t)\right)\psi.
\label{1D-NLSE}
\end{eqnarray}
The reduced two- and three-body interactions are given by, $g_{1} = \frac{m \omega_{\perp}}{2 \pi \hbar} (\frac{m}{\hbar^{2} k}) U_{1}$ and $g_{2} = \frac{m^{2}\omega_{\perp}^{2}}{3 \pi^{2}\hbar^{2}} (\frac{m}{\hbar^{2}}) U_{2}$, respectively, where, $k = 2 \pi / \lambda$. The terms inside the bracket in $g_{1}$ and $g_{2}$ are due to scaling. The time, spatial coordinate and the wavefunction are scaled as $t \rightarrow m/(\hbar k^{2})t$, $z \rightarrow z/k$ and $\psi \rightarrow \sqrt{k} \psi$, respectively. The chemical potential $\nu(t)$ and the amplitude of the lattice potential have been normalized in terms of the recoil energy $E_{R}$ \cite{choi1999bose}.

We would like to point out that most of the ultra-cold atomic systems are inhomogeneous. Due to this fact, we now consider a general potential $V_{l}(z,t) = V_{HO}(z,t) + V_{OL}(z,t)$, comprising of a harmonic trap of the form $V_{HO}(z,t) = \frac{1}{2}vM(t) z^{2}$, in addition to the time modulated lattice potential $V_{OL}(z,t) = V_{0}(t) \cos^{2}(\xi)$, with $\xi \equiv \xi(z,t) = \vartheta(t) (z - z_{0}(t))$. Here, the position of the lattice is determined by $z_{0}(t)$ and the inverse of width $\vartheta(t)$ is controlled by the frequency of the laser beam. The amplitude of the lattice potential $V_{0}(t)$, can be tuned by the intensity of the laser beam. Both $V_{0}(t)$ and $M(t)$ are scaled by the recoil energy $E_{R}$.

\section{Results: Stripe phase}

There are various methods that have been employed in order to solve the above equation, e.g., variational approximation and perturbation methods. However, in this paper, we consider a different approach, known as $F$-expansion method, which has been widely used in nonlinear fiber optics as well as in the context of BEC \cite{ali2017improved,filiz2014expansion,zhang2006improved,kruglov2003exact,lenz1993}. The F-expansion method is found to be a very useful and direct algebraic method for finding the exact solutions of nonlinear equations \cite{ali2017improved,filiz2014expansion}. The key point of this method is to transform the nonlinear system from laboratory frame to the center of mass frame (COM) through appropriate transformation, $\xi = \vartheta(t) (z - z_{0}(t))$. In other words, one looks for a travelling wave solution in the center of mass frame and transform the equation to ordinary differential equation. Without any loss of generality, we set $\hbar = m = 1$. Therefore, we assume the following ansatz in the center of mass frame:
\begin{eqnarray}
\psi(z,t) = \sqrt{\vartheta(t) \sigma(\xi)} e^{i [\Theta(\xi) + \Xi(z,t)]}.
\end{eqnarray}
Here, $\Theta(\xi)$ is a nontrivial phase, which is directly related to the velocity of the condensate and $\phi(z,t)$ is of kinematic origin,
\begin{equation}
\Xi(z,t) = \varepsilon(t) + p(t) z - \frac{1}{2} c(t) z^{2},
\end{equation}
where, the first term $\varepsilon(t)$ and $p(t)$ mimics the energy and the momentum, whereas, $c(t)$ corresponds to chirped phase. This type of chirped phase regularly arises in nonlinear fiber optics, as an acceleration induced inhomogeneity, which can balance the effect of harmonic trap.

Current conservation yields, $\partial \Theta/\partial \xi = u (1 - \frac{\sigma_{0}}{\sigma})$, where $\sigma_{0} = (\kappa^{2}_{1} - 4 \alpha \kappa_{2})/16 \kappa^{2}_{2}$ is a constant of integration and $u = \sqrt{2 \kappa^{2}_{1} - \kappa_{2} + 8 \kappa_{2} \mu}/2 \kappa_{2}$ is the velocity of the soliton for the homogeneous condensate. The COM motion obeys the following equatoin,
\begin{equation}
\frac{\partial^{2} z_{0}(t)}{\partial t^{2}} + c(t) z_{0}(t)  = \vartheta(t) (1 + u),
\label{lt.eq1}
\end{equation}
which can be written after some simple algebra in the form of a linear Schr\"odinger equation (LSE):
\begin{equation}
\frac{\partial^{2} z_{0}(t)}{\partial t^{2}} +  M(t) z_{0}(t) = 0.
\label{lt.eq}
\end{equation}
The center of mass motion of the sinusoidal excitation, also known as the Kohn mode, is found to depend solely on the frequency of the harmonic trap. It is well known that in the case of a BEC confined in a harmonic oscillator trap, the COM oscillates with the frequency of the harmonic trap. If the trap frequency varies with the time, the COM no longer oscillates with the trap frequency. In the present case, we will point out the connection between Kohn mode and the existence of the stripe phase, which we discuss in the following section.

The other parameters are found to emerge from the consistency conditions,
\numparts
\begin{eqnarray}
  \varepsilon(t) &=& \left(\mu - \frac{1}{2}\right) \int^{t}_{0}\vartheta^{2}(t) dt', \\
  \frac{\partial \vartheta(t)}{\partial t} &=&  \vartheta(t) c(t), \qquad p(t) = \vartheta(t), \\
  \frac{\partial c(t)}{\partial t} &=&   c^{2}(t) + M^{2}(t).
  \label{recatti}
\end{eqnarray}
\endnumparts
At this point, let us stop for a while and carefully analyze the above equations, which brings interesting physics as well. The eq.(\ref{recatti}), known as Reccati  equation, can be effectively mapped to LSE. Considering a change of variable, $c(t) = - \frac{\partial[\ln \varphi]}{\partial t}$, we obtain,
\begin{equation}\label{lse}
  \frac{\partial^{2} \varphi(t)}{\partial t^{2}} +  M(t) \varphi(t) = 0.
\end{equation}
This provides a correspondence between any solvable quantum mechanical potential and quasi-1D BEC. In Ref.\cite{atre2006class}, the authors mentioned that "corresponding to each solvable quantum-mechanical system, one can identify a soliton configuration", which was one of the main results of that paper. However, we rephrase and generalize the above statement, {\it correspond to each solvable quantum-mechanical system, one can identify a cnoidal wave configuration of the form $cn(\xi,k)$; in the limit $k = 1$ it corresponds to solitonic excitations \cite{atre2006class}, whereas, for $m = 0$, it mimics a sinusoidal excitations}. This allows one to investigate the dynamics of the cnoidal excitations by engineering the underlying potential through LSE.

After a lengthy algebra, the generalized NLSE reduces to the form of a generalized elliptic equation in terms of the condensate density $\sigma(\xi)$, \\
\begin{eqnarray}
\hskip-2cm
\frac{1}{4}\sigma(\xi) \frac{\partial \sigma^{2}(\xi)}{\partial \xi^{2}} - \frac{1}{8}\left(\frac{\partial \sigma(\xi)}{\partial \xi}\right)^{2} + \left(\frac{1}{2}u^{2} - \mu\right)\sigma^{2}(\xi) &+& \kappa_{1} \sigma^{3}(\xi) + \kappa_{2} \sigma^{4}(\xi) \nonumber \\ &+&
\alpha \cos^{2}(\xi) \sigma^{2}(\xi) + \frac{1}{2} \sigma_{0} = 0,
\end{eqnarray}

The reduction to a generalized elliptic equation leads to some interesting facts. We observe that the two-body interaction strength needs to be time dependent: $g_{1}(t) = \kappa_{2} \vartheta(t)$, however, the three body interaction is time independent; $g_{2} = \kappa_{2}$. The amplitude of the lattice potential has been re-scaled as, $V_{0}(t) = \alpha \vartheta^{2}(t)$. For constant $\alpha$, one finds a self-similar solution in terms of the periodic cnoidal wave $\sigma(\xi) = A + B~cn(\xi,k)$, where $k$ is the elliptic modulus parameter. However, in the present case, we are more interested in sinusoidal excitations and hence we choose $k = 0$. In this limit, the solution takes the form:
\begin{eqnarray}
\sigma(\xi) = - \frac{\kappa_{1}}{2 \kappa_{2}} \pm \sqrt{-\frac{\alpha}{\kappa_{2}}} \cos (\xi).
\end{eqnarray}

\begin{figure}[t]
\begin{center}
  \includegraphics[scale=0.478]{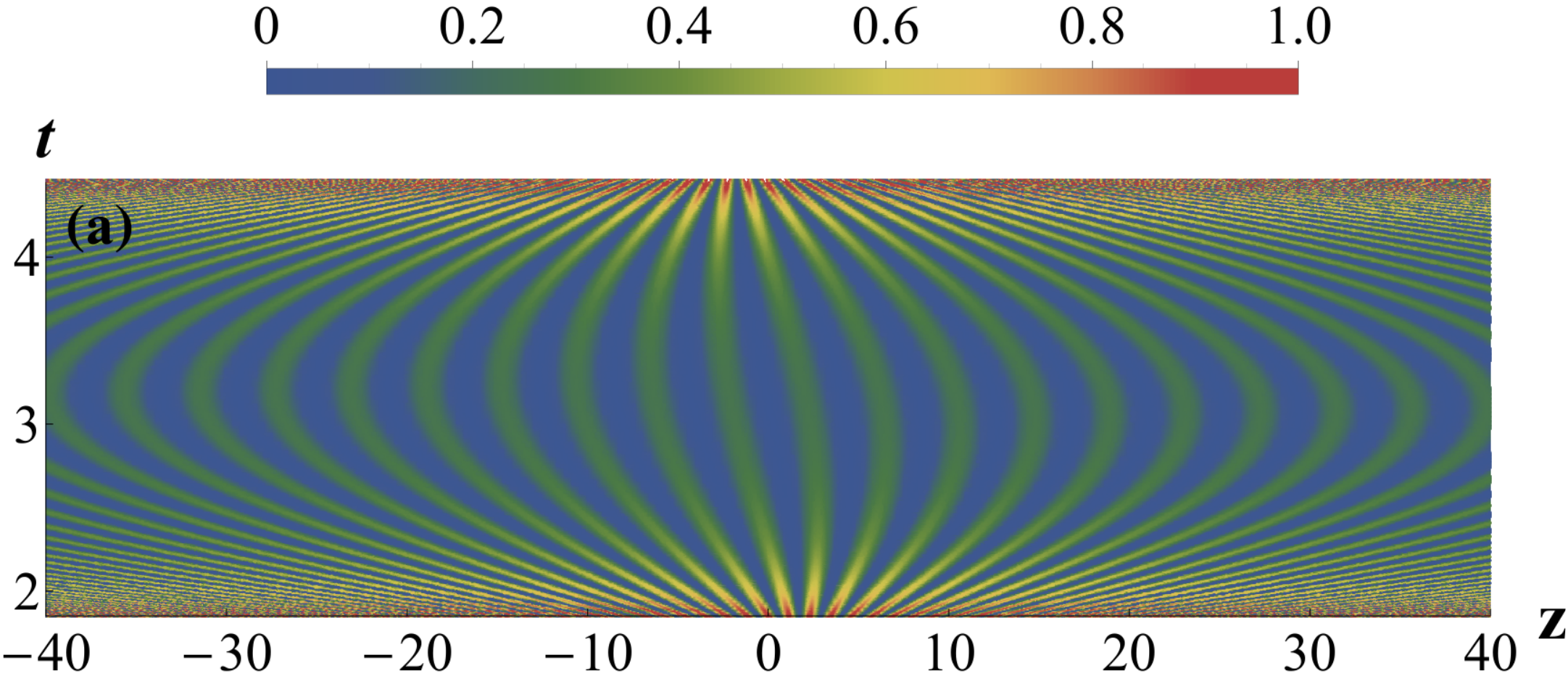}\\
  \includegraphics[scale=0.35]{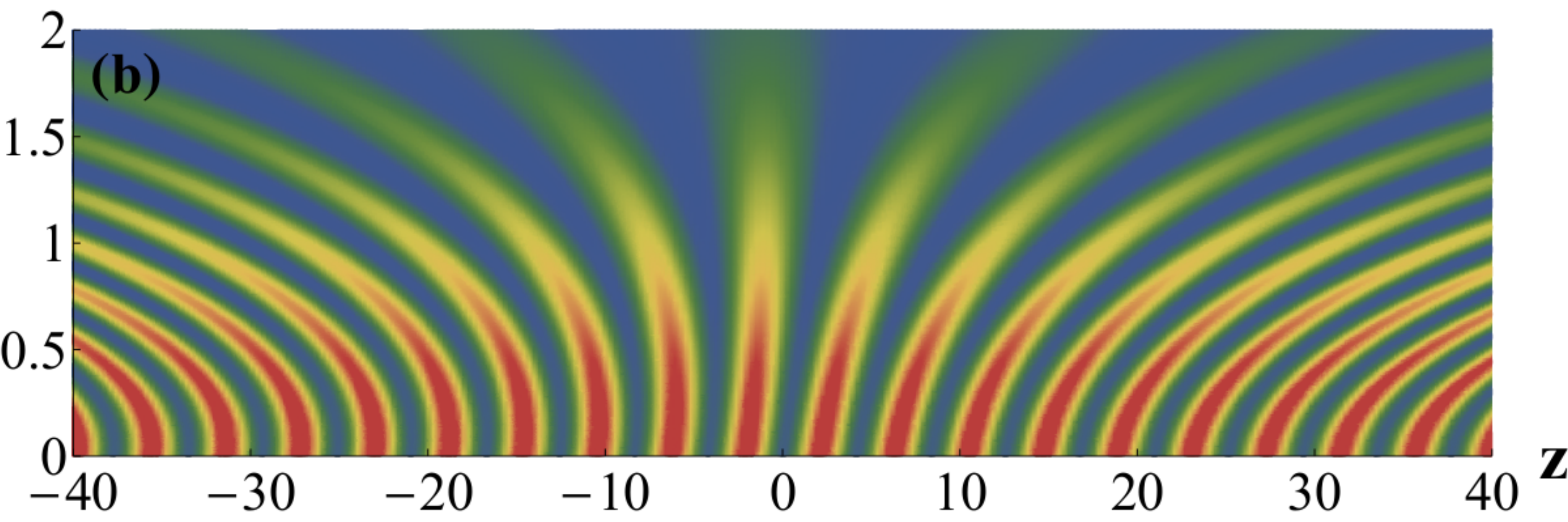}
   \caption{(Color online) The above density plots, yielding a stripe phase, are shown when the repulsive two-body and attractive three-body interactions are present. (a) depicts the stripe phase in the presence of a regular harmonic potential ($\gamma^{2} = 1$), where the chirped phase of the BEC undergoes a nonlinear compression. Existence of the stripe phase in the presence of an expulsive oscillator trap ($\gamma^{2} = -1$) is shown in (b), where the expansion of the BEC is observed. In both these cases, the other parameter values used are: $c_{0} = 1, A_{0} = 0.5, u = 0.8, \alpha = 0.15, \kappa_{1} = 0.8$ and $\kappa_{2} = -0.1$.}
   \label{fig1}
\end{center}
\end{figure}

It is worth observing that the obtained solution is quite different from the solutions existing in the literature for BEC in an optical lattice potential \cite{lin2008,zhang2007band,bronski2001}. In the present case, the periodicity of the density modulation is twice that of the lattice potential, implying the density wave character of the solution. A much deeper analysis reveals that the present solution exhibits stripe phase in certain parameter regime, where the superfluid matter is found to exist only in the finite domain. The above solution exists only when the strength of the quintic nonlinearity $\kappa_{2}$ carries opposite signature to that of both two-body interactions and the scaled amplitude of the lattice potential. Therefore, without any loss of generality, we consider the effective three-body interaction to be attractive, whereas, the two-body interaction is repulsive. Defining $\frac{\kappa_{1}}{2 \kappa_{2}} \sqrt{-\frac{\kappa_{2}}{\alpha}} = \Gamma$; the solution $\sigma(\xi)$ is found to exist for a particular parameter regime, such that $|\cos \xi| > \Gamma$. Hence, the following three cases arise, depending on the value of $\Gamma$:

\begin{description}

\item [(a)] For $\Gamma < 0$, the solution exists for all $\xi$. In this case, the three body interaction strength ($\kappa_{2}$) and $\alpha$ must have opposite signature.

\item [(b)] For $\Gamma > 1$, the solution does not exist.

\item [(c)] When $0 < \Gamma < 1$, the solution exists for $- \pi/2 < -   \xi_{j} < \xi < \xi_{j} < \pi/2$, where $\xi_{j} = \cos^{-1}\Gamma$. In this parameter regime, we observe that the superfluid matter wave can only be found in the finite domain, in alternate lattice sites, indicating the existence of a stripe phase. This is a unique characteristic of the present system, where such ordered phase appears purely due to the interplay between quintic nonlinearity and lattice potential. The translational symmetry of the condensate spatial profile is spontaneously broken due to the presence of the lattice. Furthermore, the presence of quintic nonlinearity induces a double periodicity in the density profile. Unlike the pure cubic nonlinear case, where periodicity of the density modulation is same as the optical lattice, in the present case, the periodicity of the density modulation is twice that of the lattice potential. This phenomenon is analogous to the density wave, where the superfluid matter exists in alternative lattice sites.
\end{description}

\section{Discussions}

\subsection{Non-inertial frame of reference}

\begin{figure}[t]
\begin{center}
    \includegraphics[scale = 0.57]{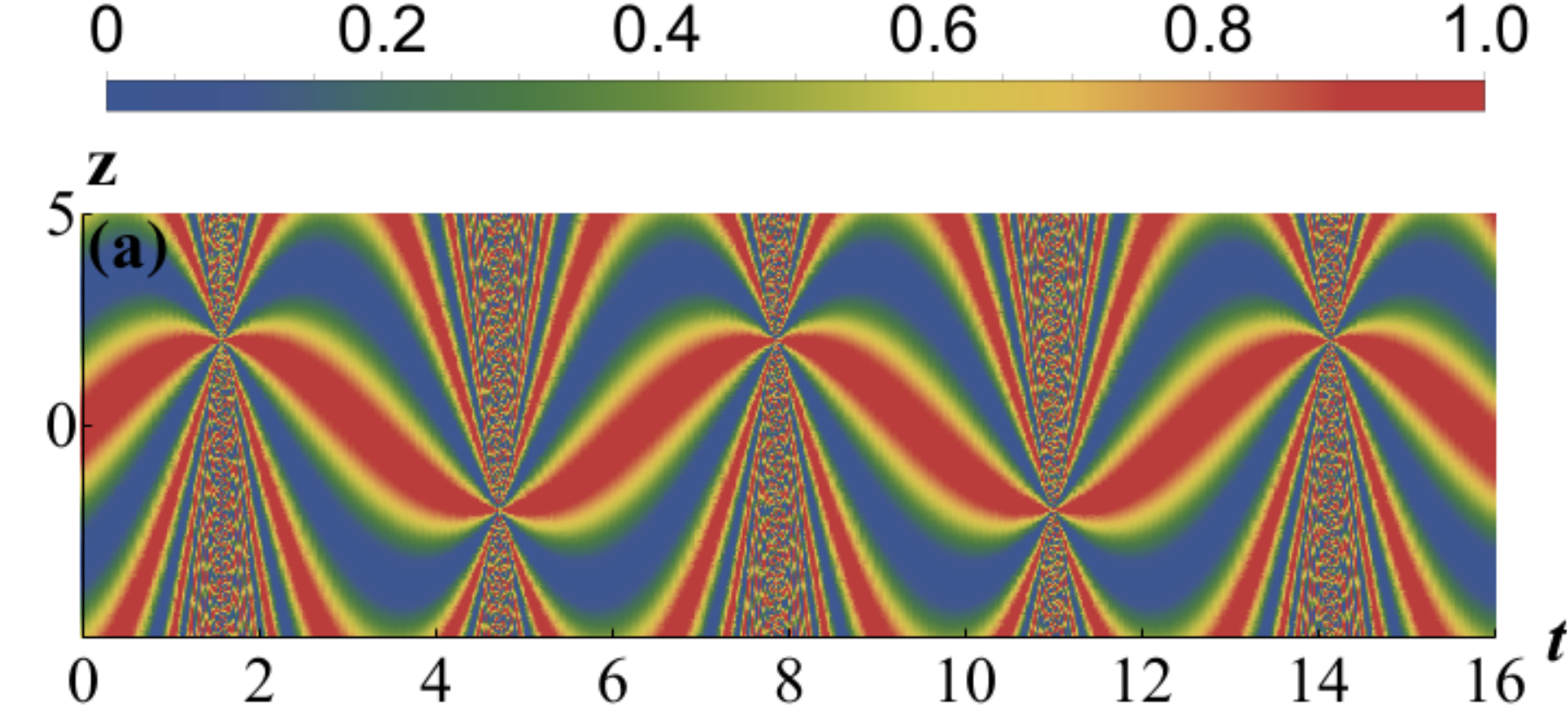}\\
    \includegraphics[scale = 0.41]{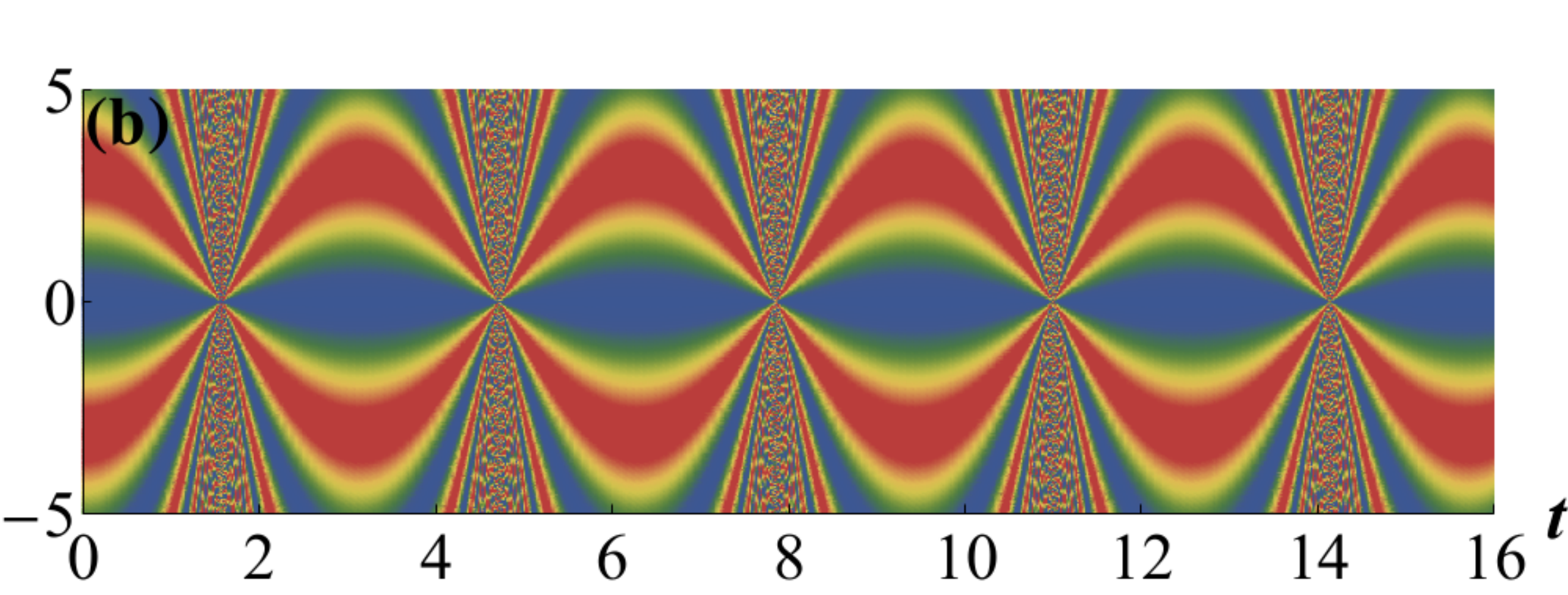}
\caption{(Color online) The behavior of the supercurrent is shown in presence of an regular harmonic trap. (a) correspond to $u = 1$ and and fig.(b) is obtained for $-1$. In both the cases, the supercurrent is found to change its direction at the point of nonlinear compression. The parameter values used are same as Fig.(\ref{fig1}).  }
\label{fig2}
\end{center}
\end{figure}

In this section, we discuss the existence of stripe phase in different scenarios, when the harmonic trap is present along with a time modulated optical lattice potential. In the present case, we consider $0 < \Gamma <1 $, where a stripe phase exists due to the interplay between the modulated lattice potential and the interactions, and not due to the harmonic trap. In the following section, we show the existence of the stripe phases when the harmonic trap is switched off. In presence of the harmonic trap, we observe that the COM moves with finite acceleration, and hence such frame of reference is referred as a non-inertial frame. We initially consider the strength of the harmonic trap to be a constant, e.g., $M(t) = \gamma^{2}$. This represents a regular harmonic trap and the corresponding position in the center of mass frame is: $l(t) = l_{0} \sin (\gamma t)$. The associated chirped phase and the inverse of width follow: $c(t) = \gamma \tan(\gamma t)$ and $\vartheta(t) = A_{0} Sec(\gamma t)$. The stripe ordering of the condensate is shown in Fig.(\ref{fig1}a). The edges near $t \sim 1.8$ and $t \sim 4.8$ are the high-density regime, also known as nonlinear compression of the BEC. Such high density arises from the condition when the harmonic trapping frequency is in resonance with the chirped pulses. The coupling of the interactions and the lattice potential spontaneously breaks the translational symmetry of the BEC, leading to a stripe phase. In the presence of a regular harmonic trap, the COM motion undergoes an oscillation with the trap frequency and thereby satisfies Kohn theorem. However, when the harmonic trapping frequency depends on time, the center of mass motion does not oscillate with the trapping frequency and the stripe phase disappears. This is due to the fact that in presence of a time dependent harmonic trap e.g., $M(t) = e^{- \gamma t}$, the chirped phase is described by the Bessel function and consequently destroys the underlying periodicity of the system and hence, the stripe phase disappears. Furthermore, the supercurrent takes the following form,
\begin{equation}
J(t) = u (\sigma - \sigma_{0}).
\end{equation}
The behavior of supercurrent is shown in Fig.(\ref{fig2}). Fig.(\ref{fig2}a) corresponds to $u = 1$. This induces a positive effective net phase in the COM motion (see r.h.s. of Eq.(\ref{lt.eq1})) and the point of nonlinear compression shifts away from $z = 0$. Fig. (\ref{fig2}b) depicts the variations of supercurrent for $u = -1$, where one can see that the point of nonlinear compression lies at $z = 0$, since no net phase is accumulated in COM motion. Furthermore, it is interesting to observe that the supercurrent changes its direction at the point of nonlinear compression, as can be seen in both Fig.(\ref{fig2}a) and (\ref{fig2}b).

Inspired by the work \cite{liang2005}, we consider another scenario, where the harmonic trap is considered as an inverted trap, i.e., $M(t) = -\gamma^{2}$, also known as an expulsive oscillator. Such inverted trap has been considered in order to observe the nonlinear resonances or nonlinear compression of solitonic excitations. In this case, the center of mass motion follows a localized profile, $l(t) = l_{0} \sinh(\alpha t)$. Corresponding chirped pulse and the inverse of the width, both are found to be localized, as expected, $c(t) = \gamma~\tanh(\gamma t)$ and $\vartheta(t) = A_{0}~\textrm{sech}(\gamma t)$. In this case, we also observe the stripe phase, which exists only in the finite domain, with a spatially broken symmetry. The corresponding evolution of the density is shown in Fig.(\ref{fig1}b). As the time increases, the amplitude of the propagating wave decreases, making the BEC spread out. Hence, the competition between time-dependent lattice amplitude and the oscillator trap deforms the simple periodic structure of the matter waves.

\begin{figure}[t]
\begin{center}
  \includegraphics[scale=0.48]{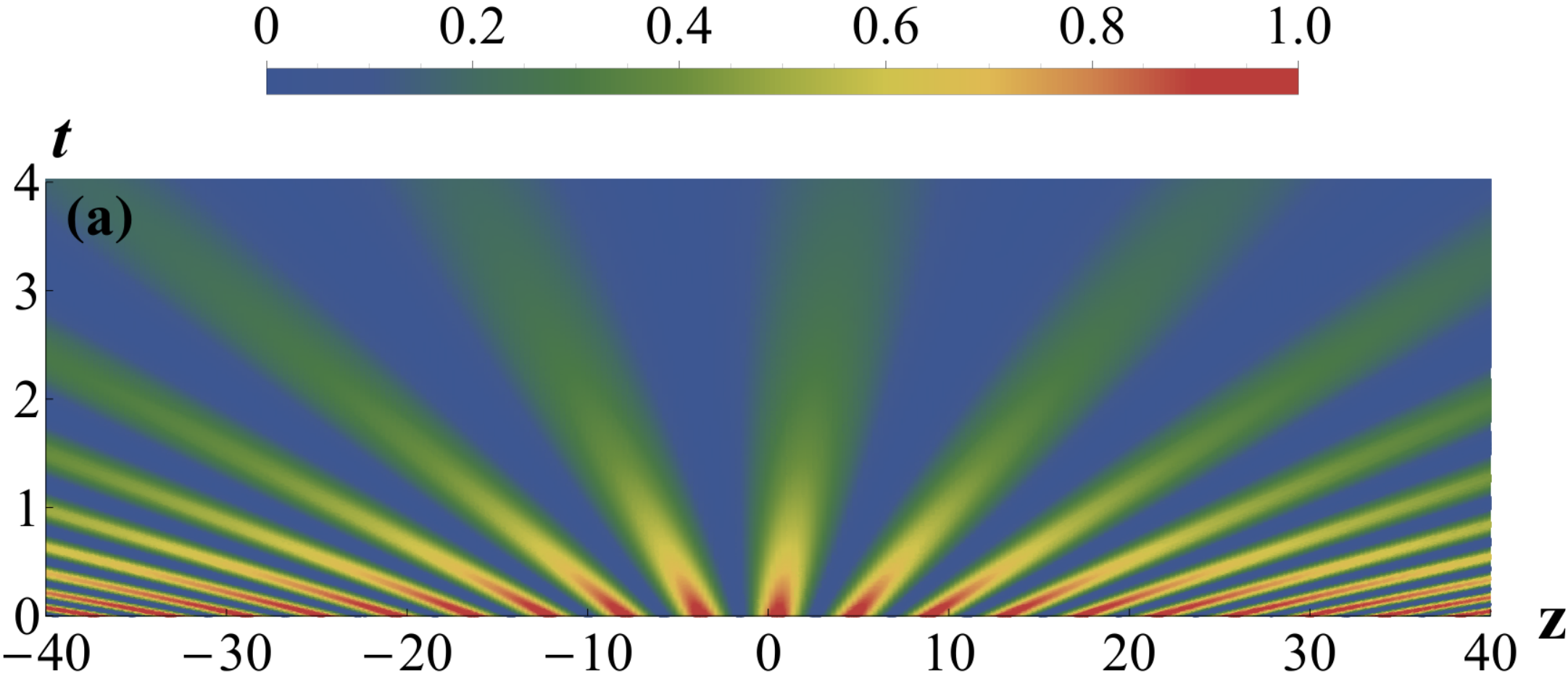} \\
  \includegraphics[scale=0.35]{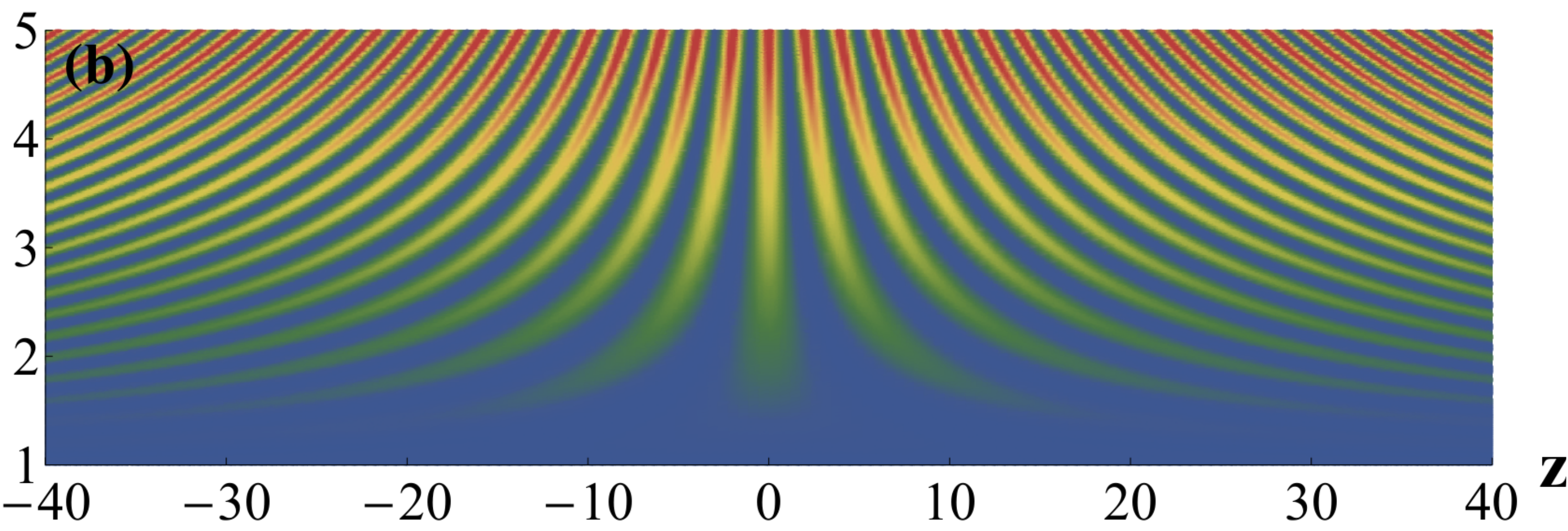}
   \caption{(Color online) The above density plots are obtained when the harmonic trap is switched off. (a) depicts the stripe phase when the COM is moving with a constant velocity. When the COM is static, the stripe phase still exists and is shown in (b). The parameter values used are same as Fig.(\ref{fig1}), except $\gamma^{2} = 0$.}
   \label{fig3}
\end{center}
\end{figure}

\subsection{Inertial frame of reference}

In this section, we investigate the existence of the superstripes, when the harmonic trap is switched off, i.e., $M(t) = 0$. Therefore, from Eq.(\ref{lt.eq}), one can clearly see that the COM motion follows a linear path, and hence, its acceleration is zero. Therefore, the center of mass is moving with a constant velocity and such frame of reference is known as the inertial frame. However, in this case, interestingly we find that the chirped pulse is still finite even in the absence of harmonic trap. The form of the COM motion is, $l(t) = v_{0} t + l_{0}$, where $v_{0}$ and $l_{0}$ are constants. The lattice is moving with a constant velocity $v_{0}$. The associated chirped phase and the inverse of the width are, respectively, found to be: $c(t) = - 1/(t + c_{0})$ and $\vartheta(t) = A_{0}/(t + c_{0})$. Both the chirped pulse and the inverse of width are inversely proportional to the time. Fig. (\ref{fig3}a) depicts the corresponding striped behavior of the condensate when the lattice is moving with a constant $v_{0}$. Similar to the expulsive oscillator case, here also we find that during evolution, the amplitude of the condensate decreases, implying the expansion if the BEC. However, due to the absence of the harmonic trap, the distortion in the striped phase is absent in this case.

For the sake of completeness, we also analyze the system, when the optical lattice is static. This implies, the position of the COM, $l(t) = 0$ or takes a constant value. The chirped phase and the inverse of width in this case, turn out to be of the form: $c(t) = c_0/(c_0 t - 1)$ and $\vartheta(t) = A_{0} (c_{0}t - 1)$. The stripe phase is still observable in this case, which is shown in Fig. (\ref{fig3}b). The amplitude of the condensate is increasing during the evolution of the BEC, which mimics that the condensate undergoes a nonlinear compression. This phenomenon of the nonlinear compression is analogous to the effective pulse compression in nonlinear optical fiber and analogous to the findings of Moores \cite{moores1996nonlinear}. Therefore, it is evident that the self-similar sinusoidal excitations can be controlled by means of chirp management, leading to the controlled nonlinear compression and amplification.

\begin{figure}[h]
\begin{center}
 \includegraphics[scale=0.3]{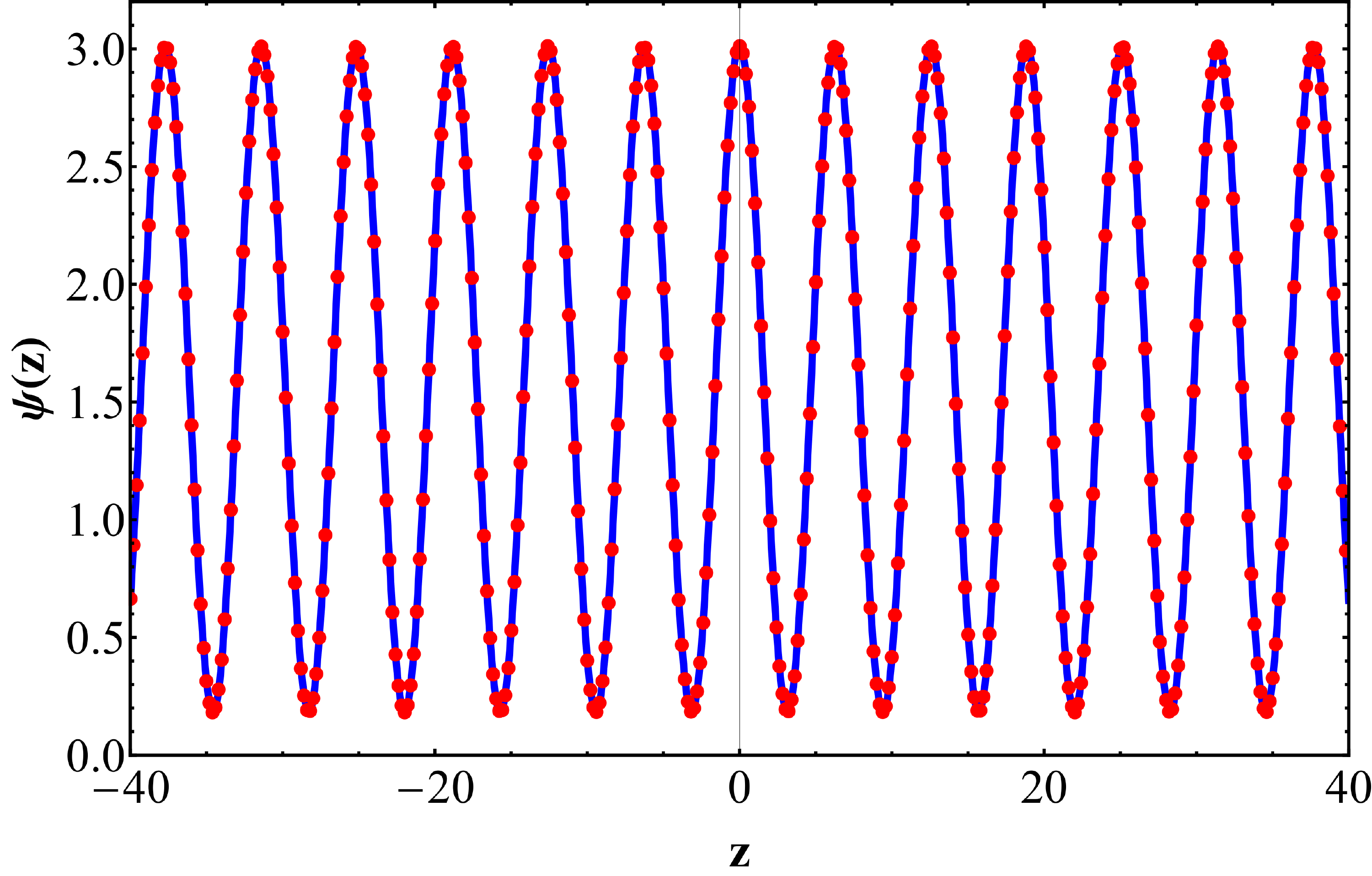}
\caption{(Color online)  The figure shows the results obtained from linear stability analysis. The blue, solid curve corresponds to the analytically obtained solution. The red dotted curve depicts the the evolution of the perturbed solution, which coincides with the blue curve, mimicking the obtained solution is stable. The parameter values are same as fig.(\ref{fig1}), except $\kappa_{2} = -0.25$.}
  \label{fig8}
\end{center}
\end{figure}

\section{Linear stability analysis}

Till now, we have focused on the stripe phase that exists through the spontaneously broken spatial symmetry of the BEC. In this section, we perform a linear stability analysis of the obtained phases. Applying a small perturbation ($\delta \psi \ll 1$) to the obtained solution, $\psi(z,t) = \psi(x) + \delta \psi(z,t)$. Therefore, one can write the equation of motion in terms of the perturbation $\delta \psi$, after substituting the above ansatz in eq.(\ref{1D-NLSE}).
\begin{eqnarray}\label{per}
\hskip-1.5cm
i \frac{\partial \delta\psi}{\partial t} = \left(- \frac{1}{2} \frac{\partial{^2} }{\partial z^{2}} + V_{l} - \nu(t)\right)\delta\psi &+& g_{1}(2 |\psi|^{2} \delta \psi + \psi^{2} \delta \psi^{*}) \nonumber \\ &+& g_{2}(3 |\psi|^{4} \delta \psi + 2 |\psi|^{2} \psi^{2} \delta \psi^{*} ).
\end{eqnarray}
where, we set $\hbar = m = 1$. Fig.(\ref{fig8}) illustrates the obtained exact analytical solution and the perturbed solutions obtained numerically by solving eq.(\ref{per}). Due to the presence of a harmonic trap and periodic lattice potentials, we use mixed boundary conditions, $\delta\psi(0,t) - i \delta_{z} \delta\psi(0,t) = 0$ and $\delta\psi(L,t) - i \delta_{z} \delta\psi(L,t) = 0$. In our numerical simulation, we use the split-step Crank-Nicholson (CN) scheme, where $dx = 0.01$ and $dt = 0.00005$, and thereby satisfying the CN scheme $dz^{2}/dt > 1$. It can be clearly seen from Fig.(\ref{fig8}) that the obtained solution is linearly stable. Furthermore, we obtained the following quantity \cite{soto1991stability,akhmediev1985excitation,das2013realization}
\begin{equation}
  \Omega = \frac{\ln\{ \textrm{Re}[\delta\psi(z, t + \delta t)] \} - \ln\{ \textrm{Re}[\delta\psi(z, t)] \}}{dt}
\end{equation}
at each step of our simulation, which in the limit of $t \rightarrow \infty$ and $\delta t \rightarrow 0$ mimics the largest eigenvalue of the perturbation mode, obtained by solving eq.(\ref{per}). Explicit numerical simulation shows that the eigenvalues correspond to the perturbation mode, obtained by using eq.(\ref{per}) are negative and hence, the solution is found to be stable.

\begin{figure}[t]
 \includegraphics[scale=0.3]{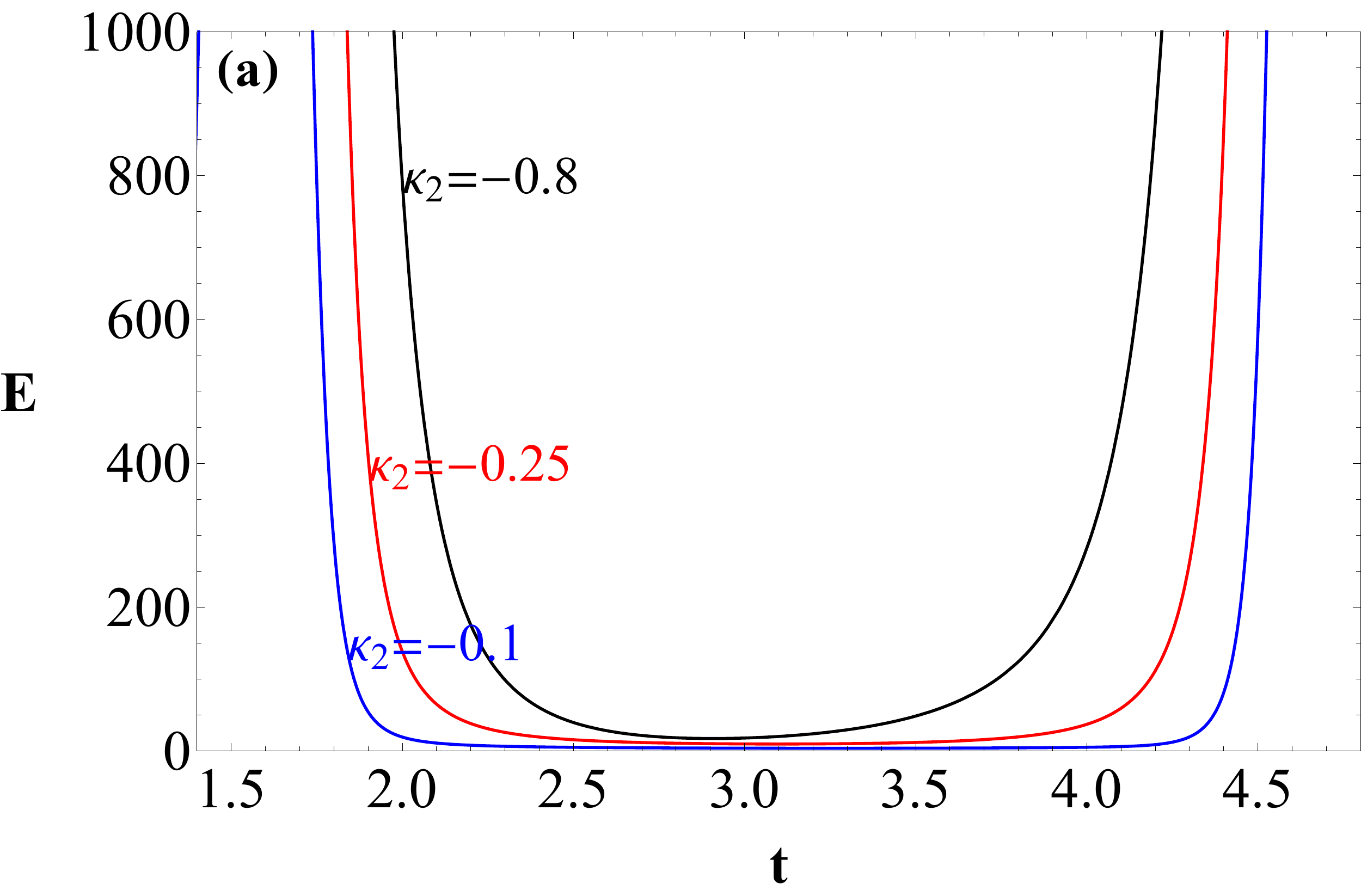}
 \includegraphics[scale=0.3]{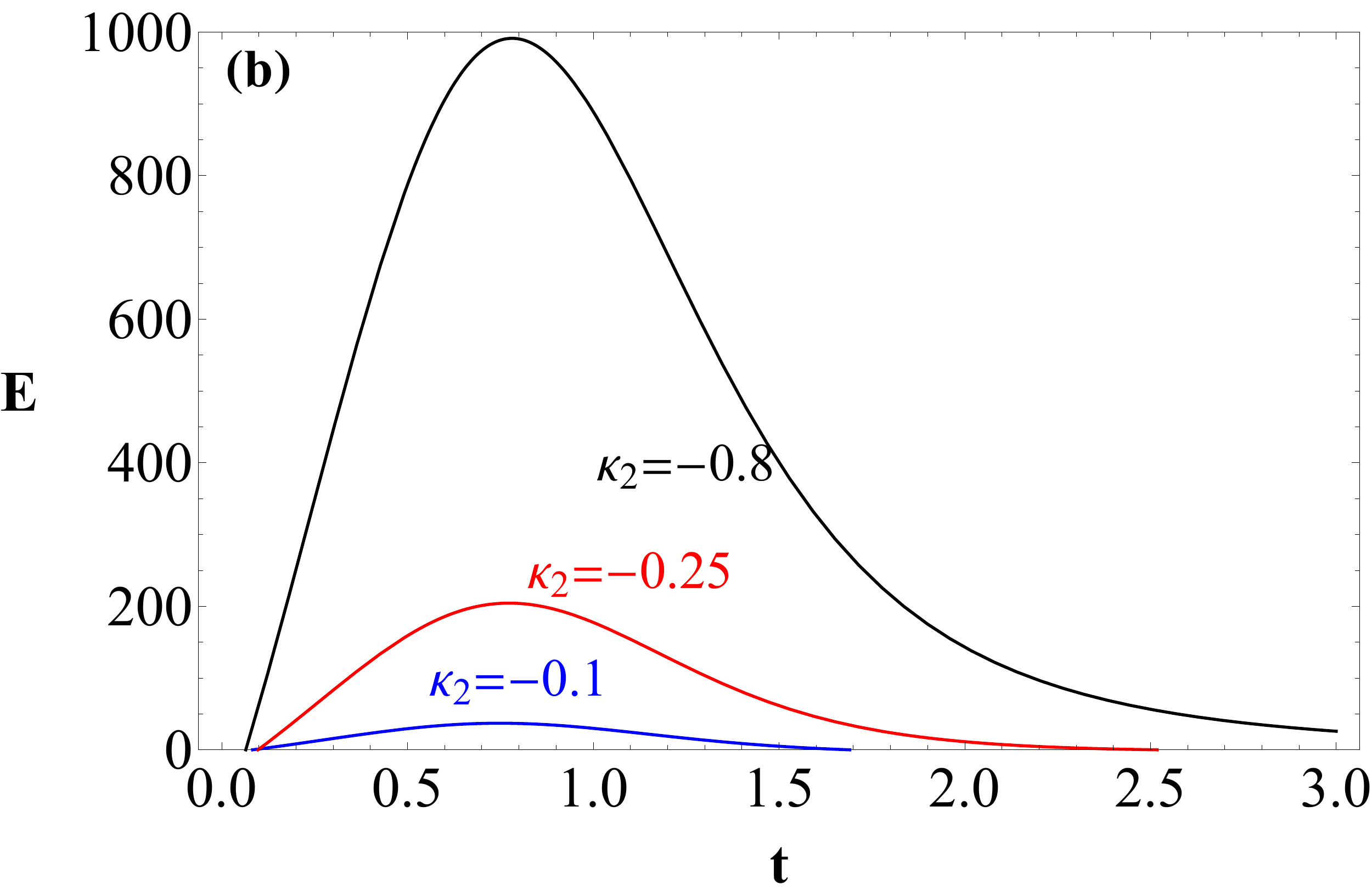}
\caption{(Color online) The energy spectrum is shown in the combined presence of the harmonic trap and lattice potential. (a) The behavior of the energy in the presence of a regular harmonic trap, where the nonlinear resonances appear at certain time domain, which exactly matches with the density profile fig.(\ref{fig1}a). (b) In case of expulsive oscillator trap, the energy initially increases and then decreases, mimicking the expansion of BEC. The parameter values used are same as fig.(\ref{fig1}).}
  \label{fig4}
\end{figure}

\section{Energy spectrum}

\begin{figure}[t]
 \includegraphics[scale=0.3]{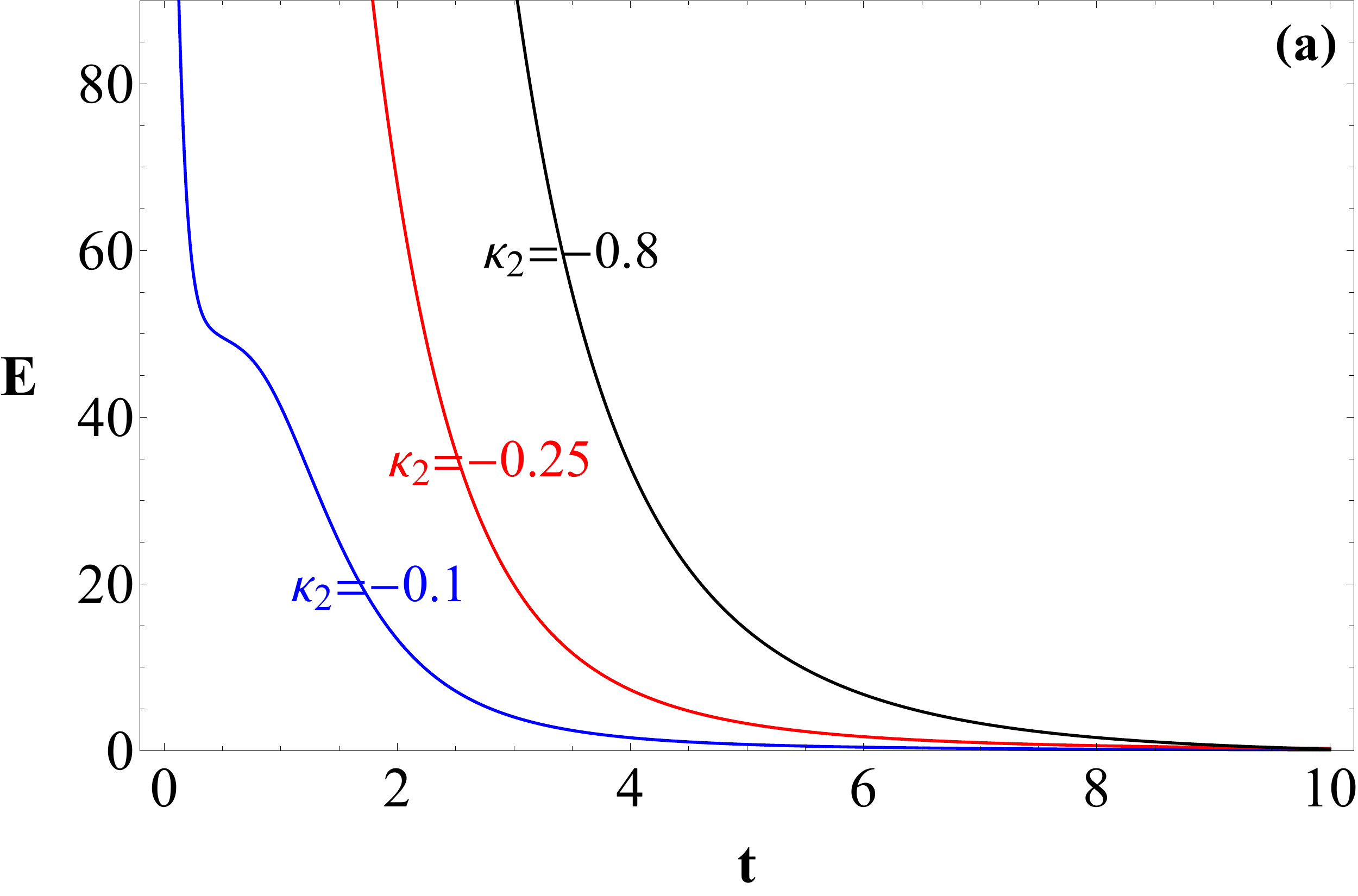}
 \includegraphics[scale=0.3]{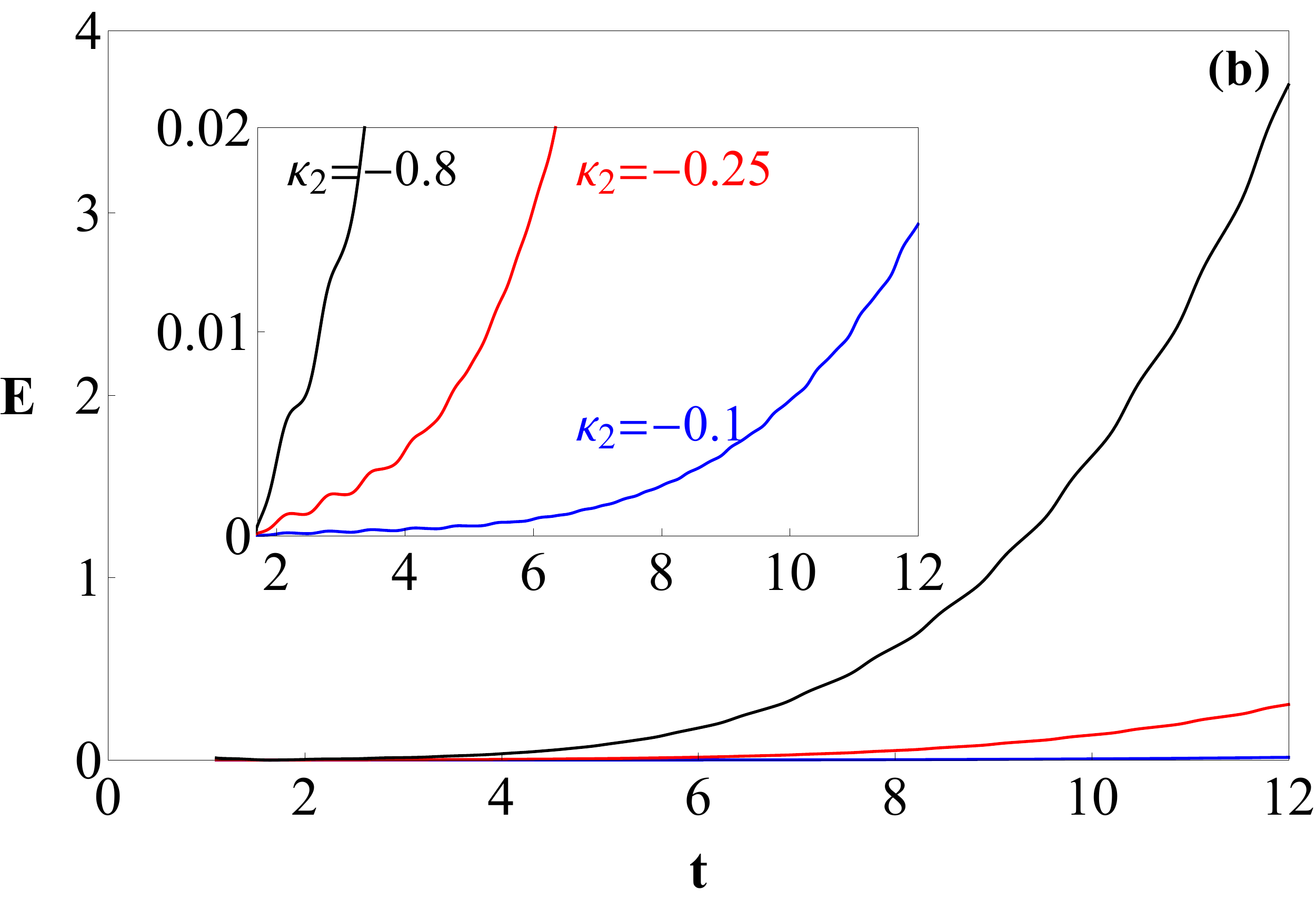}
\caption{(Color online) The energy spectrum is shown when the harmonic trap is set to zero. (a) The behavior of the energy when the COM is moving with a constant velocity. The energy is found to decrease in this case, yielding the expansion of BEC. (b) A resonant behavior is observed when the COM is static. The increase in energy mimics the nonlinear compression of the condensate. The parameter values are same as fig.(\ref{fig1}).}
  \label{fig5}
\end{figure}

The analytical calculation of energy with accelerating optical lattice is nontrivial. The energy profile is shown in Fig.(\ref{fig4}) and (\ref{fig5}), both in the presence and absence of the harmonic trap. The temporal behavior of energy in this parametrically driven system is studied. In particular, the effect of three-body interaction strength on energy is investigated. In this scenario, the positive-definiteness of the density modulation allows for repulsive two-body and attractive three-body interactions. Fig.(\ref{fig4}) depicts the time evolution of the energy spectrum in presence of the combined presence of the lattice potential and harmonic trap. In presence of regular harmonic trap ($M = \gamma^{2}$), we observe the nonlinear resonance due to strong nonlinear compression of the condensate. This can be clearly seen from Fig.(\ref{fig4}a), where the energy spectrum is shown for three different values of $\kappa_{2}$. For, $\kappa_{2} = -1$, the density of the condensate is found to be high around $t \sim 1.8$ and $t \sim 4.8$ (Fig.(\ref{fig1}a)), subsequently, at the same position, the energy increases abruptly, which gives rise to the nonlinear compression of the condensate. Similar feature is seen for other values of the quintic nonlinearity, $\kappa_{2} = -0.25$ (red curve) and $-0.8$ (black curve).

For the sake of completeness. we have checked that the potential energy does not exhibit such resonant behavior. The contribution of the supercurrent to the kinetic energy is solely responsible for this phenomenon. It is interesting to mention that in an experiment of BEC in an optical lattice, similar resonant behavior has been observed \cite{fabbri2009}. Fig.(\ref{fig4}b) depicts the behavior of the energy for expulsive harmonic trap ($M = \gamma^{2}$). The energy is found to increase initially and then decreases. In this case, the periodic excitations initially move towards to center of the trap, gains its energy. However, for an expulsive trap, the maximum position is unstable and hence, the excitations follow the downstream path and subsequently the corresponding energy decreases. The dynamics become completely different in the absence of the  harmonic trap, where the lattice potential plays a dominant role. When the lattice is moving with a constant velocity, the energy is decreasing, due to the expansion of the condensate, as is seen in Fig.(\ref{fig5}a). We observe a resonance behavior when the lattice is static since the BEC undergoes a nonlinear compression. Such resonance is analogous to the effective pulse  compression in nonlinear optical fiber, observed by Moore \cite{moores1996nonlinear}.

\section{Conclusion}

In summary, we have presented here a detailed study of the parametrically forced Bose-Einstein condensate in the combined presence of the lattice and harmonic potential, where both cubic and quintic nonlinearities are present. We observe a spontaneously broken spatial symmetry of the condensate, mimicking the existence of a stripe phase in this system. We show that such stripe phase exists due to the interplay between lattice potential and the quintic nonlinearity. The presence of a lattice potential induces a periodicity in the system. However, due to the presence of the quintic nonlinearity, the periodicity of the density modulation is found to be twice that of the lattice potential. This imposes a restriction on the density modulations and hence, the superfluid matter waves are found to exist only in a finite domain, yielding a stripe phase. A linear stability analysis shows that the obtain solution is stable.  Due to the presence of a harmonic trap, we consider a chirped phase in the system, which leads to an efficient nonlinear compression of the condensate in certain parameter regime. In order to gain a better understanding of the nonlinear compression, we compute the energy of the system. The corresponding energy spectrum clearly shows the regimes of the nonlinear resonances, which leads to strong nonlinear compression of the condensate. We discussed that such resonance behavior arises when the harmonic trapping frequency is in resonance with the chirped pulses.

\section*{Acknowledgements}

The author thanks Prasanta K. Panigrahi, Anirban Pathak and Ayan Khan for many useful discussions. The author also acknowledges the financial support provided by the Department of Atomic Energy (DAE), Government of India, under the project no. RP01378.

\section*{References}

\bibliographystyle{iopart-num}
\bibliography{ms}

\providecommand{\newblock}{}
\begin{thebibliography}{10}
\expandafter\ifx\csname url\endcsname\relax
  \def\url#1{{\tt #1}}\fi
\expandafter\ifx\csname urlprefix\endcsname\relax\def\urlprefix{URL }\fi
\providecommand{\eprint}[2][]{\url{#2}}

\bibitem{morsch2006}
Morsch O and Oberthaler M 2006 {\em Rev. Mod. Phys.\/} {\bf 78}(1) 179--215

\bibitem{beinke2017}
Beinke R, Klaiman S, Cederbaum L~S, Streltsov A~I and Alon O~E 2017 {\em Phys.
  Rev. A\/} {\bf 95}(6) 063602

\bibitem{lewenstein2007ultracold}
Lewenstein M, Sanpera A, Ahufinger V, Damski B, Sen A and Sen U 2007 {\em
  Advances in Physics\/} {\bf 56} 243--379

\bibitem{jaksch2003creation}
Jaksch D and Zoller P 2003 {\em New Journal of Physics\/} {\bf 5} 56

\bibitem{nakano2017bose}
Nakano E, Yabu H and Iida K 2017 {\em Phys. Rev. A\/} {\bf 95}(2) 023626
  \urlprefix\url{https://link.aps.org/doi/10.1103/PhysRevA.95.023626}

\bibitem{chikkatur2000}
Chikkatur A~P, G\"orlitz A, Stamper-Kurn D~M, Inouye S, Gupta S and Ketterle W
  2000 {\em Phys. Rev. Lett.\/} {\bf 85}(3) 483--486

\bibitem{chang2008crystallization}
Chang D, Gritsev V, Morigi G, Vuleti{\'c} V, Lukin M and Demler E 2008 {\em
  Nature Physics\/} {\bf 4} 884--889

\bibitem{zohar2015quantum}
Zohar E, Cirac J~I and Reznik B 2015 {\em Reports on Progress in Physics\/}
  {\bf 79} 014401

\bibitem{bloch2005ultracold}
Bloch I 2005 {\em Nat. Phys.\/} {\bf 1} 23

\bibitem{bloch2005quantum}
Bloch I 2005 {\em J. Phys. B: At. Mol. Opt. Phys.\/} {\bf 38} S629

\bibitem{chen2008}
Chen G, Wang X, Liang J~Q and Wang Z~D 2008 {\em Phys. Rev. A\/} {\bf 78}(2)
  023634

\bibitem{bloch2008}
Bloch I, Dalibard J and Zwerger W 2008 {\em Rev. Mod. Phys.\/} {\bf 80} 885

\bibitem{kartashov2011}
Kartashov Y~V, Malomed B~A and Torner L 2011 {\em Rev. Mod. Phys.\/} {\bf
  83}(1) 247--305

\bibitem{lin2008}
Lin G~D, Zhang W and Duan L~M 2008 {\em Phys. Rev. A\/} {\bf 77} 043626

\bibitem{zhang2007band}
Zhang A~X and Xue J~K 2007 {\em Phys. Rev. A\/} {\bf 75}(1) 013624

\bibitem{smerzi2002}
Smerzi A, Trombettoni A, Kevrekidis P~G and Bishop A~R 2002 {\em Phys. Rev.
  Lett.\/} {\bf 89}(17) 170402

\bibitem{bronski2001}
Bronski J~C, Carr L~D, Deconinck B and Kutz J~N 2001 {\em Phys. Rev. Lett.\/}
  {\bf 86}(8) 1402--1405

\bibitem{leonard2017supersolid}
L{\'e}onard J, Morales A, Zupancic P, Esslinger T and Donner T 2017 {\em
  Nature\/} {\bf 543} 87--90

\bibitem{baumann2010dicke}
Baumann K, Guerlin C, Brennecke F and Esslinger T 2010 {\em Nature\/} {\bf 464}
  1301--1306

\bibitem{fallani2004}
Fallani L, De~Sarlo L, Lye J~E, Modugno M, Saers R, Fort C and Inguscio M 2004
  {\em Phys. Rev. Lett.\/} {\bf 93} 140406

\bibitem{cataliotti2003superfluid}
Cataliotti F~S, Fallani L, Ferlaino F, Fort C, Maddaloni P and Inguscio M 2003
  {\em New J. of Phys.\/} {\bf 5} 71

\bibitem{greiner2002quantum}
Greiner M, Mandel O, Esslinger T, H{\"a}nsch T~W and Bloch I 2002 {\em
  Nature\/} {\bf 415} 39

\bibitem{mivehvar2015}
Mivehvar F and Feder D~L 2015 {\em Phys. Rev. A\/} {\bf 92}(2) 023611

\bibitem{martone2014}
Martone G~I, Li Y and Stringari S 2014 {\em Phys. Rev. A\/} {\bf 90}(4) 041604

\bibitem{li2013}
Li Y, Martone G~I, Pitaevskii L~P and Stringari S 2013 {\em Phys. Rev. Lett.\/}
  {\bf 110}(23) 235302

\bibitem{wang2010}
Wang C, Gao C, Jian C~M and Zhai H 2010 {\em Phys. Rev. Lett.\/} {\bf 105}(16)
  160403

\bibitem{emery1999stripe}
Emery V, Kivelson S and Tranquada J 1999 {\em Proceedings of the National
  Academy of Sciences\/} {\bf 96} 8814--8817

\bibitem{wu2006quantum}
Wu C, Liu W~V, Moore J and Sarma S~D 2006 {\em Phys. Rev. Lett.\/} {\bf 97}(19)
  190406 \urlprefix\url{https://link.aps.org/doi/10.1103/PhysRevLett.97.190406}

\bibitem{demler2002quantum}
Demler E, Wang D~W, Sarma S~D and Halperin B 2002 {\em Solid state
  communications\/} {\bf 123} 243--250

\bibitem{pattabhiraman2017formation}
Pattabhiraman H and Dijkstra M 2017 {\em Soft Matter\/}

\bibitem{dunjko2001}
Dunjko V, Lorent V and Olshanii M 2001 {\em Phys. Rev. Lett.\/} {\bf 86}(24)
  5413--5416

\bibitem{paredes2004tonks}
Paredes B, Widera A, Murg V, Mandel O, F{\"o}lling S, Cirac I, Shlyapnikov G~V,
  H{\"a}nsch T~W and Bloch I 2004 {\em Nature\/} {\bf 429} 277--281

\bibitem{choi2015monopole}
Choi S, Dunjko V, Zhang Z~D and Olshanii M 2015 {\em Phys. Rev. Lett.\/} {\bf
  115}(11) 115302

\bibitem{adhikari2017}
Adhikari S~K 2017 {\em Laser Physics Letters\/} {\bf 14} 065402

\bibitem{anatoly2009}
Kamchatnov A~M and Salerno M 2009 {\em Journal of Physics B: Atomic, Molecular
  and Optical Physics\/} {\bf 42} 185303

\bibitem{das2009loss}
Das P, Vyas M and Panigrahi P~K 2009 {\em J. Phys. B: At. Mol. Opt. Phys.\/}
  {\bf 42} 245304

\bibitem{abdullaev2015faraday}
Abdullaev F~K, Gammal A and Tomio L 2015 {\em Journal of Physics B: Atomic,
  Molecular and Optical Physics\/} {\bf 49} 025302

\bibitem{lahaye2008}
Lahaye T, Metz J, Fr\"ohlich B, Koch T, Meister M, Griesmaier A, Pfau T, Saito
  H, Kawaguchi Y and Ueda M 2008 {\em Phys. Rev. Lett.\/} {\bf 101}(8) 080401

\bibitem{moores1996nonlinear}
Moores J~D 1996 {\em Optics letters\/} {\bf 21} 555--557

\bibitem{agrawal2007nonlinear}
Agrawal G~P 2007 {\em Nonlinear fiber optics\/} (Academic press)

\bibitem{abdullaev2005gap}
Abdullaev F~K and Salerno M 2005 {\em Phys. Rev. A\/} {\bf 72} 033617

\bibitem{salasnich2007matter}
Salasnich L, Malomed B~A and Toigo F 2007 {\em Phys. Rev. A\/} {\bf 76} 063614

\bibitem{salasnich2002effective}
Salasnich L, Parola A and Reatto L 2002 {\em Phys. Rev. A\/} {\bf 65} 043614

\bibitem{utpal2010complex}
Roy U, Atre R, Sudheesh C, Kumar C~N and Panigrahi P~K 2010 {\em J. Phys. B:
  At. Mol. Opt. Phys.\/} {\bf 43} 025003

\bibitem{choi1999bose}
Choi D~I and Niu Q 1999 {\em Phys. Rev. Lett.\/} {\bf 82} 2022

\bibitem{ali2017improved}
Ali~Akbar M and Ali N~H~M 2017 {\em Cogent Mathematics\/} {\bf 4} 1282577

\bibitem{filiz2014expansion}
Filiz A, Ekici M and Sonmezoglu A 2014 {\em The Scientific World Journal\/}
  {\bf 2014}

\bibitem{zhang2006improved}
Zhang J~L, Wang M~L, Wang Y~M and Fang Z~D 2006 {\em Physics Letters A\/} {\bf
  350} 103--109

\bibitem{kruglov2003exact}
Kruglov V~I, Peacock A~C and Harvey J~D 2003 {\em Phys. Rev. Lett.\/} {\bf 90}
  113902

\bibitem{lenz1993}
Lenz G, Meystre P and Wright E~M 1993 {\em Phys. Rev. Lett.\/} {\bf 71} 3271

\bibitem{atre2006class}
Atre R, Panigrahi P~K and Agarwal G~S 2006 {\em Phys. Rev. E\/} {\bf 73} 056611

\bibitem{liang2005}
Liang Z~X, Zhang Z~D and Liu W~M 2005 {\em Phys. Rev. Lett.\/} {\bf 94} 050402

\bibitem{soto1991stability}
Soto-Crespo J, Heatley D, Wright E~M and Akhmediev N 1991 {\em Physical Review
  A\/} {\bf 44} 636

\bibitem{akhmediev1985excitation}
Akhmediev N, Korneev V and KUZMENKO I 1985 {\em Zhurnal Eksperimental'noi i
  Teoreticheskoi Fiziki\/} {\bf 88} 107--115

\bibitem{das2013realization}
Das P, Noh C and Angelakis D~G 2013 {\em EPL (Europhysics Letters)\/} {\bf 103}
  34001

\bibitem{fabbri2009}
Fabbri N, Cl\'ement D, Fallani L, Fort C, Modugno M, van~der Stam K~M~R and
  Inguscio M 2009 {\em Phys. Rev. A\/} {\bf 79} 043623

\end{thebibliography}

\end{document}